%% file: MTGR.tex
\documentclass[sigconf]{acmart}
\usepackage{multirow}
\usepackage{balance}
\newcommand{\method}{\text{MTGR}}
\newcommand{\user}{{\mathbf{U}}}
\newcommand{\seq}{\mathbf{S}}
\newcommand{\rt}{\overrightarrow{\mathbf{R}}}
\newcommand{\iteminseq}{\mathbf{s}}
\newcommand{\itemfeature}{\mathbf{{I}}}
\newcommand{\cross}{\mathbf{{C}}}
\newcommand{\Emb}{\mathbf{Emb}}
\newcommand{\Feat}{\mathbf{Feat}}
\newcommand{\uni}{\text{model}}
\newcommand{\X}{\mathbf{X}}
\newcommand{\K}{\mathbf{K}}
\newcommand{\Q}{\mathbf{Q}}
\newcommand{\V}{\mathbf{V}}
\newcommand{\U}{\mathbf{U}}
\AtBeginDocument{%
  }

\copyrightyear{2025}
\acmYear{2025}
\setcopyright{acmlicensed}\acmConference[CIKM '25]{Proceedings of the 34th ACM International Conference on Information and Knowledge Management}{November 10--14, 2025}{Seoul, Republic of Korea}
\acmBooktitle{Proceedings of the 34th ACM International Conference on Information and Knowledge Management (CIKM '25), November 10--14, 2025, Seoul, Republic of Korea}
\acmDOI{10.1145/3746252.3761565}
\acmISBN{979-8-4007-2040-6/2025/11}
\begin{document}

\title{\method{}: Industrial-Scale Generative Recommendation Framework in Meituan}

% \author{XXXXXXXX}
% \affiliation{%
%   \institution{Meituan}
%   \city{Beijing}
%   % \state{Ohio}
%   \country{China}
% }
% \author{Ruidong Han,Bin Yin,Shangyu Chen,Chi Ma,He Jiang,Mincong Huang,Xiaoguang Li,Chunzhen Jing,Fei Jiang,Lei Yu,Xiang Li,Wei Lin}
% \affiliation{%
%   \institution{Meituan}
%   \city{Beijing}

\author{Ruidong Han, Bin Yin, Shangyu Chen, He Jiang, Fei Jiang, Xiang Li, Chi Ma, Mincong Huang, Xiaoguang Li, Chunzhen Jing, Yueming Han, Menglei Zhou, Lei Yu, Chuan Liu, Wei Lin}
\authornote{Corresponding author.}
\affiliation{%
\{hanruidong, yinbin05, chenshangyu03, jianghe06, jiangfei05, lixiang245, machi04, huangmincong, lixiaoguang12, jingchunzhen, hanyueming02, zhoumenglei, yulei37, liuchuan11, linwei31\}@meituan.com
\\
  \institution{Meituan}
  \city{Beijing}
  \country{China}}

\renewcommand{\shortauthors}{Ruidong Han, Bin Yin, Shangyu Chen et al.}

\begin{abstract}
\input{Contents/abstract}
\end{abstract}







\maketitle

\section{Introduction}
\input{Contents/introduction}

\section{Related Works}
\input{Contents/related-works}

\section{Preliminary}
\input{Contents/preliminary}

\section{Data Rearrangement and Architecture of \method{}}
\input{Contents/method}

\section{Training System}
\input{Contents/training}

\section{Experiments}
\input{Contents/experiments}

\section{Conclusion}
\input{Contents/conclusion}

\clearpage
% \section{Reference}
\balance
\bibliographystyle{ACM-Reference-Format}
\bibliography{reference}

\clearpage
\onecolumn
% \section{Appendix}
% \input{Contents/appendix}

\end{document}

%% file: Contents/abstract.tex
Scaling law has been extensively validated in many domains such as natural language processing and computer vision. 
In the recommendation system, recent work has adopted generative recommendations to achieve scalability, but their generative approaches require abandoning the carefully constructed cross features of traditional recommendation models. 
We found that this approach significantly degrades model performance, and scaling up cannot compensate for it at all.
In this paper, we propose $\method{}$ (Meituan Generative Recommendation) to address this issue. 
$\method{}$ is modeling based on the HSTU \cite{zhai2024actions} architecture and can retain the original deep learning recommendation model (DLRM) features, including cross features. 
Additionally, MTGR achieves training and inference acceleration through user-level compression to ensure efficient scaling.
We also propose Group-Layer Normalization (GLN) to enhance the performance of encoding within different semantic spaces and the dynamic masking strategy to avoid information leakage. 
We further optimize the training frameworks, enabling support for our models with 10 to 100 times computational complexity compared to the DLRM, without significant cost increases.
$\method{}$ achieved $65$x FLOPs for single-sample forward inference compared to the DLRM model, resulting in the largest gain in nearly two years both offline and online. This breakthrough was successfully deployed on Meituan, the world's largest food delivery platform, where it has been handling the main traffic.

%% file: Contents/introduction.tex
Scaling law have been proven to apply to most deep learning tasks, including language models\cite{kaplan2020scaling}, computer vision\cite{zhai2022scaling,peebles2023scalable}, and information retrieval\cite{fang2024scaling}.
Our work focuses on scaling up the ranking model in industrial recommendation systems effectively.
% Recommendation system predicts user preference for candidate items, given the features of the user and candidates.
Under the requirements of high QPS (Queries Per Second) and low latency in industrial recommendation systems, the scaling of the model is usually limited by both the training cost and inference time.
Currently, the study on scaling ranking models can be divided into two types: Deep Learning Recommendation Model (DLRM) and Generative Recommendation Model (GRM). 
DLRM models individual user-item pairs to learn the probability of interest for ranking and scales up by developing more complex mappings. 
GRM organizes data by token like natural language and performs next token prediction through a transformer architecture.

In industrial recommendation systems, DLRM has been used for nearly a decade, typically with inputs that include a large number of well-designed handcrafted features such as cross features \footnote{Cross features measure the interactions of multiple raw features, like the user's historical click-through rate for the target candidate} to improve model performance. 
However, DLRM has two significant drawbacks when it comes to scaling:
1) with exponential growth of user behavior, traditional DLRM cannot efficiently process entire user behaviors, often resorting to sequence retrieval or designing low-complexity modules for learning, which limits the model's learning capability \cite{pi2020search, chen2021end}.
2) scaling based on DLRM incurs approximately linear costs in training and inference with the number of candidates, making the expenses unbearably high.

For GRM, recent studies have pointed out its excellent scalability \cite{zhai2024actions, deng2025onerec}. We identify two key factors:
1) GRM directly models the complete chain of user behavior, which compresses multiple samples of exposure under the same user into one. This significantly reduces computational redundancy, while allowing end-to-end encoding of longer sequence compared to DLRM; 
2) GRM adopts a transformer architecture with efficient attention computation\cite{zhai2024actions, dao2022flashattention}, enabling the model's training and inference to meet the requirements of industrial recommendation systems. 
% However, GRM is not aware of the specific item being predicted during generation and therefore cannot use any cross features.
However, GRM heavily relies on next token prediction to model a complete user behavior sequence, which requires removing cross features between candidates and the user.
We found that excluding cross features severely damages the model's performance, and this degeneration cannot be compensated by scaling up at all.

How can we build a ranking model that utilizes cross features to ensure effectiveness while also possessing the scalability of GRM?
To address this problem, we propose the Meituan Generative Recommendation ($\method{}$).
Compared to traditional DLRM and GRM, $\method{}$ utilizes the advantages and discards disadvantage of the methods. 
$\method{}$ maintains inputs feature consistent with traditional DLRM, including cross features.
Specially, $\method{}$ reorganizes features by converting user and candidate features into different tokens, leading to a token sequence for efficient model scaling. % The representation of candidate token is utilized for learning using a discriminative loss. 
Then $\method{}$ incorporates the cross feature in the candidate tokens and learns with a discriminative loss. 

% $\method{}$ employs a user sample aggregation like GRM for training acceleration, while it does not model the entire chain of user behavior. Instead, it splits into user tokens and item tokens similar to DLRM and uses a discriminative loss to predict item tokens. Therefore, it can maintain feature inputs consistent with traditional DLRM, including cross features.

MTGR employs the similar HSTU (Hierarchical Sequential Transduction Units) architecture as used in \cite{zhai2024actions} for modeling. In HSTU, we propose Group-layer Normalization(GLN) to normalize different types of token separately, enabling better modeling of multiple heterogeneous information simultaneously. 
In addition, we propose a dynamic masking strategy, where 
full-attention, auto-regressive and visibility only to itself are used to ensure performance and avoid information leakage.
% and to enhance the perceptual range between different tokens and avoid information leakage. 

Unlike the commonly used TensorFlow in the industry, MTGR training framework is built based on TorchRec\cite{ivchenko2022torchrec} and optimized for computational efficiency.
Specifically, to handle the real-time insert/delete of sparse embedding entries, we employ dynamic hash tables to replace static tables. To improve efficiency, we conduct dynamic sequence balancing to address the computation load imbalances among GPUs and adopt embedding ID de-duplication alongside automatic table merging to accelerate embedding lookup. We also incorporate implementation optimization including mixed precision training and operator fusion. Compared to TorchRec, our optimized framework improves training throughput by $1.6$x – $2.4$x while achieving good scalability when running over $100$ GPUs.

%checkpoint resuming, mixed precision training, gradient accumulation, and operator fusion. Compared to TorchRec, our optimized framework improves training throughput by $1.6x$ – $2.4x$ while achieving good scalability when running over $100$ GPUs.

We validate the scalability of $\method{}$ on a small-scale dataset. Then, we design three models of varying sizes, which are trained using following over six months of data, to verify the scaling law for both offline and online performance.
The large version, compared to the DLRM baseline optimized over years, demonstrates $65$x FLOPs per sample for forward computation and increases the conversion volumes by $1.22\%$ and the CTR (Click-Through Rate) by $1.31\%$ in our scenarios. 
Meanwhile, the training cost remains unchanged and the inference cost is reduced by 12\%. $\method{}$-large has been deployed in the Meituan take-away recommendation system, serving hundreds of millions of users.

In summary, our contributions are summarized as follows:
\begin{itemize}
\item $\method{}$ combines the advantages of DLRM and GRM, retaining all the features of DLRM, including the cross feature, while demonstrating excellent scalability as GRM.

\item We propose Group-Layer Normalization and dynamic masking strategies to achieve better performance.

\item We perform systematic optimizations on the TorchRec-based $\method{}$ training framework to enhance training performance.

\item Both offline and online experiments were conducted to demonstrate the power-law relationship between the performance of MTGR and computational complexity, and its superiority compared to DLRM.

\end{itemize}

%% file: Contents/related-works.tex
\subsection{Deep Learning Recommendation Model}
A classic DLRM structure typically includes many inputs such as context (e.g. time, location), user profile (e.g. gender, age), user behavior sequences, and target item with many cross features. 
Two particularly important modules in ranking model are behavior sequence processing and feature interactions learning. 
The behavior sequence module\cite{zhou2018deep, pi2020search, si2024twin} usually employs target attention mechanisms to capture the similarity between the historical behavior of the users and the item to be estimated. 
The feature interactions module\cite{lian2018xdeepfm, tang2020progressive,wang2021dcn,wang2024home} is designed to capture interactions among different features including user and item to produce the final prediction. 

\subsection{Scaling up Recommendation Model}
Based on the different scaling modules in DLRM, there are two distinct approaches.  One approach is the scaling cross module, i.e. scaling up the feature interactions module that integrates user and item information. \cite{zhang2024wukong} introduces a stackable Wukong layer for scaling up. \cite{guo2023embedding} employ a multi-embedding strategy to address the embedding collapse, thereby enhancing the model's scalability. 
The other approach is the scaling user module, where only the user part is scaled up, making this approach more inference-friendly. \cite{zhang2024scaling, han2024enhancing} reduce online inference costs by scaling only user representations and caching or broadcasting them to different items that will be estimated.  \cite{shin2023scaling, yan2025unlocking} design a pre-training method for user representations, demonstrating scalability in downstream tasks. 

The counterpart to DLRM is GRM. \cite{zhai2024actions} validate the scaling law through HSTU with scaling up to trillion-level parameters. \cite{deng2025onerec} use semantic coding to replace traditional ID representations, combining DPO optimization with a transformer-based framework to replace the cascaded learning framework with a unified generative model.

%% file: Contents/preliminary.tex
\subsection{Data Arrangement}\label{sec:data-arrangment}
% Traditionally when the rank module in the commercial recommendation system is requested by a user, it retrieves user's features and counterparts of candidates from the recall, leading to multiple inference samples for scoring and ranking. 
% One request to the recommendation system leads to input samples with size of $K$, which is also the number of candidates.
Traditionally, for a user and the corresponding $K$ candidates, the $i$-th sample for the user and $i$-th candidate pair can be represented as $\mathbb{D}_i = [\user{}, \overrightarrow{\seq{}}, \rt{}, \cross{}_i, \itemfeature{}_i]$. % Features used are categorized into 
Specifically, $\user{} = [\user{}^1, ..., \user{}^{N_{\user{}}}]$ represents the user's profile feature ($\user{}^i$), such as age, gender, etc. 
Each feature $\user{}^i$ is a scalar, and $N_{\user{}}$ represents the number of features used.
$\overrightarrow{\seq{}} = [\seq{}^1, ..., \seq{}^{N_{\seq{}}}]$ contains the sequence of items that have historically been interacted with by the user. Each element in $\seq{}^i = [\iteminseq{}^1, ..., \iteminseq{}^{N_{\iteminseq{}}} ]$ represents an item, it is comprised by selected features ($\iteminseq{}^i$) such as the item's ID, tag, average CTR on the item and etc.
Similar to $\overrightarrow{\seq{}}$, $\rt{}$ records the closet interaction to current request within a few hours or a day. $\rt$ represents user's real-time actions and preference.
It shares the same features with $\overrightarrow\seq{}$.
$\cross{} = [\cross{}^1,...,\cross{}^{N_{\cross{}}}]$ comprises the cross features between the user and the candidates. 
$\itemfeature{} = [\itemfeature{}^1,...,\itemfeature{}^{N_{\itemfeature{}}}]$ contains the features used for candidates, such as item's ID, tag, and brand. $\itemfeature{}$ is candidate-dependent and shared for different users.

\subsection{Ranking Model in Recommendation Systems}
\begin{figure}
    \centering
    \includegraphics[width=\linewidth]{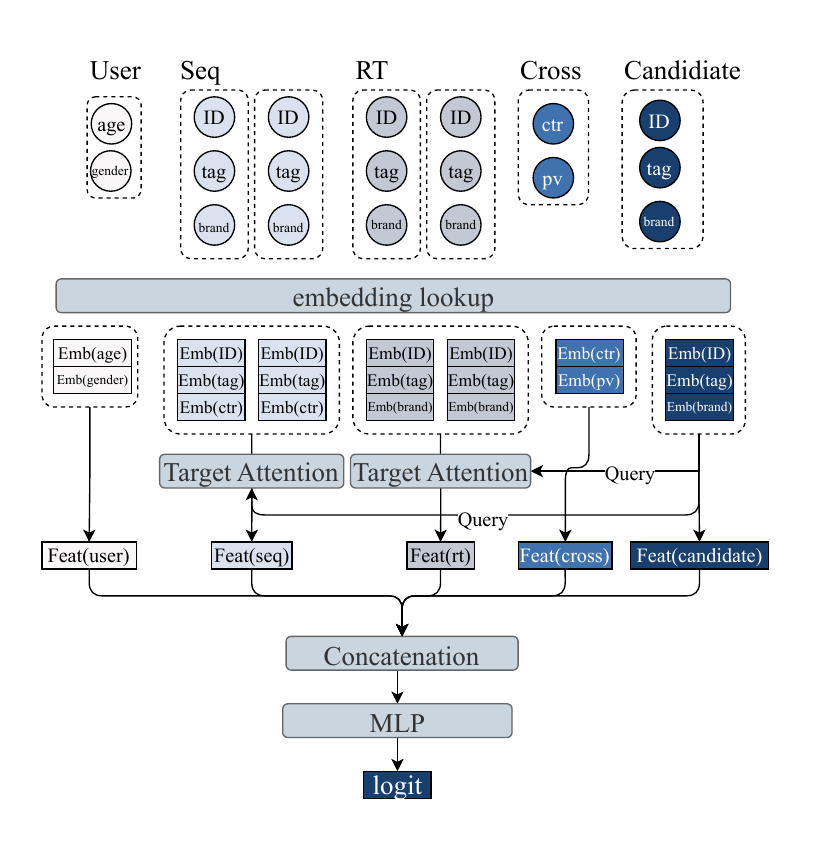}
    \captionsetup{font={small}}
    \caption{Data Arrangement and workflow of traditional ranking model. 
    An example with simplified features is demonstrated: $\user{}$ contains `age' and `gender'; $\overrightarrow\seq{}$ and $\rt{}$ is comprised by 2 item respectively, each item possesses `ID', `tag' and `brand'; $\cross{}$ contains `ctr' and `pv', presenting user's historical CTR and number of exposures to the candidate, which uses `ID', `tag' and `brand'.
    }
    \vspace{-0.5em}
    \label{fig:traditional-recsys}
\end{figure}
With input samples $\mathbb{D}$, traditional recommendation system processes the samples independently.
Specially, it firstly embeds the features in $\mathbb{D}$ and converts the samples to a dense representation.
Formally, the features in $\user{}$, $\cross{}$, $\itemfeature{}$ are embedded and concatenated to $\Emb{}_{\user{}} \in \mathbb{R}^{K \times d_{\user{}}}$, $\Emb{}_{\cross{}} \in \mathbb{R}^{K \times d_{\cross{}}}$ and $\Emb{}_{\itemfeature{}} \in \mathbb{R}^{K \times d_{\itemfeature{}}}$, respectively.
For features in $\overrightarrow{\seq{}}$ and $\rt{}$ \footnote{In the following description, since $\overrightarrow{\seq{}}$ and $\rt{}$ are processed in a similar way, only $\overrightarrow{\seq{}}$'s modeling is described for clarity.}, each item($\seq{}^i$)'s feature are similarly embedded and concatenated to $\Emb_{\seq{}^i} \in \mathbb{R}^{d_{\iteminseq{}}}$, items in $\overrightarrow{\seq{}}$ are concatenated in other dimension, leading to $\Emb_{\overrightarrow\seq{}} \in \mathbb{R}^{N_{\overrightarrow \seq{}} \times d_{\iteminseq{}}}$ \footnote{$N_{*}$ represents the sequence length of $*$}.

To extract user interest between the historical interacted items and candidates, target attention normally be used with target as query and sequence feature as key/value. 
Formally, 
\begin{eqnarray}\label{eq:target-attention}
    \Feat{}_{\overrightarrow\seq{}} = \text{Attention}(\Emb{}_{\itemfeature{}}, \Emb_{\overrightarrow\seq{}}, \Emb_{\overrightarrow\seq{}}) \in \mathbb{R}^{K \times d_{\seq{}}}
\end{eqnarray}
Eq.\ref{eq:target-attention} aggregates $\overrightarrow\seq{}$ according to $\itemfeature{}$. 
Finally, embedded and processed features from $\mathbb{D}$ are concatenated and represented as:
\begin{eqnarray}
    \Feat{}_{\mathbb{D}} = [\Emb{}_{\user{}}, \Feat{}_{\overrightarrow\seq{}}, \Feat{}_{\rt{}}, \Emb{}_{\cross{}}, \Emb_{\iteminseq{}}] \in \mathbb{R}^{K \times (d_{\user{}} + d_{\seq{}} + d_{\cross{}} + d_{\itemfeature{}})}
\end{eqnarray}
$\Feat{}_{\mathbb{D}}$ is further fed to a multiple layer perceptron (MLP) to output logit for each sample. The logit is used for learning in training and for ranking when inference.

Fig.\ref{fig:traditional-recsys} shows a simplified data arrangement and workflow under the traditional ranking model.
% Specially, $\user{}$ contain feature `age' and `ctr' (the user's ctr over all exposure). $\overrightarrow \seq{}$ contains 2 items, each of which is comprised by feature `ID', `tag' and `ctr' (the item's ctr over all exposure). In $\cross{}$, we use `ctr', `pv' as example. In $\itemfeature{}$, `ID', `tag' and `ctr' (the item's ctr to the user) are utilized. 
These features are firstly embedded, the leading embeddings are processed in different methods. 
% Embedding for $\user{}$ is fed into user module, which is typically a MLP or self-attention module.
% Since different user contains different length of sequence, the items' embedding in the sequence are attended by embedding of targets, where items serve as key and value while targets serve as query, leading to feature of sequence.
Finally, the processed features are concatenated and processed by MLP for feature interaction. The final logit is generate for each candidates.

\subsection{Scaling Dilemma in Recommendation Systems}
Model scaling has been a common method for performance improvement in ranking. Generally, model scaling aims at scaling parameters in user module and cross module.
The user module processes user feature including sequence features and generates user-dependent representation. Scaling user module brings in better representation for user. Besides, since the user is shared and inferred once for all candidates, a large inference cost in the user module does not lead to a burdensome system overload. However, scaling user module alone does not contribute to feature interaction in user and item directly.

On the contrary, another trend of method aims at scaling cross module, i.e., the feature interaction MLP after feature concatenation. These methods enhance ranking ability by paying more attention to interaction between user and candidates. However, since cross module is inferred for each candidates, computation increasing is scaled linearly with number of candidates. Cross module scaling brings in unacceptable system latency.

The scaling dilemma in traditional recommendation system necessitates a new scaling method, leading to efficient feature interaction between user and candidates, while at the same time brings sub-linear inference cost with number of candidates.
\method{} innovates the scaling approach in recommendation system by data rearrangement and corresponding architecture optimization.

%% file: Contents/method.tex
\subsection{User Sample Aggregation for Training and Inference Efficiency}
\begin{figure}
    \centering
    \includegraphics[width=\linewidth]{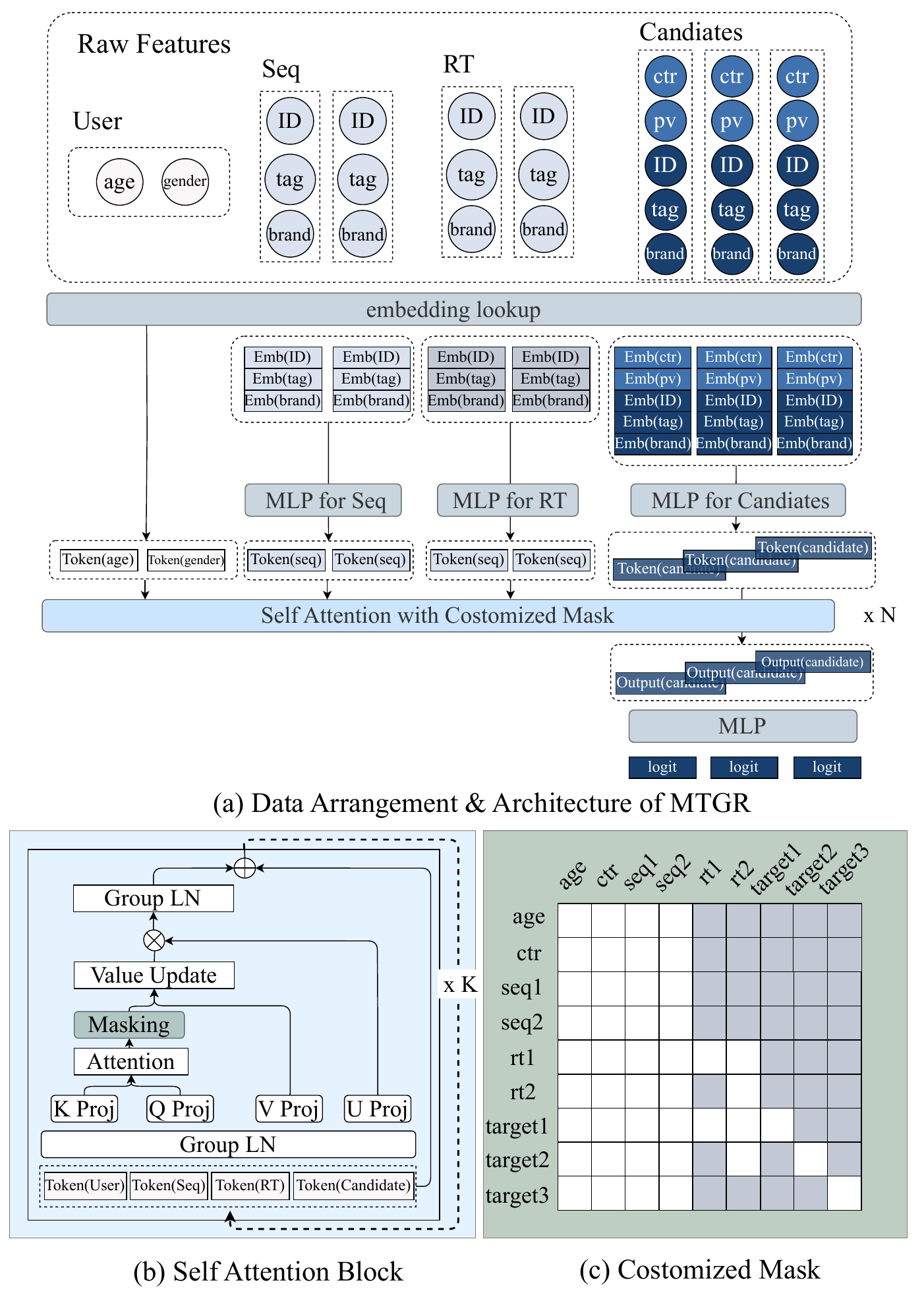}
    \captionsetup{font={small}}
    \caption{Data Arrangement and architecture of $\method{}$. 
    (a) represents data arrangement and overall workflow of $\method{}$: 3 candidates' features are aggregated with the counterpart of one user. The features are embedded and converted to token by MLP, forming a series of input sequence for self-attention with mask. The representation of the tokens of candidates are used for logit via another MLP module. 
    (b) Detailed description of self-attention module of $L$ layers: input sequence is firstly processed by a group layer norm and 4 projections (Q/K/V/U). Self-attention is conducted with a customized mask. After value update, a dot-producted is performed on updated V and projected U. Finally, the updated value is again group-wise layer normed and added with a residual connection. % Similar to \cite{zhai2022scaling}, FFN is not utilized in $\method{}$.
    (c) Customized mask to avoid information leakage, `rt's and `target's are ordered from recent to past: user feature ($\user{}$, $\overrightarrow \seq{}$) is visible to all the tokens; `rt's is visible to other tokens appear later; `target's is visible to itself only.
    }
    \vspace{-0.5em}
    \label{fig:gr-workflow}
\end{figure}
Compared with feature categorization in Sec.\ref{sec:data-arrangment}, for $i$-th sample in candidates, \method{} organizes features into $\mathbb{D}_i = [\user{}, \overrightarrow\seq{}, \rt{}, [\cross{}_i, \itemfeature{}_i]]$. 
Specially, the cross feature $\cross{}$ is arranged as part of the item feature of candidates.
In \method{}, candidates are aggregated by user in a specific window for training and by request for inference.  
% the user-dependent feature remains unchanged for different candidates. 
% For training samples, user's exposures during the aggregation window are grouped. During inference, the candidates retrieved in the request are aggregated. 
Since the aggregation is performed by the same user, the aggregated sample can use the same user representation ($\user{}$, $\overrightarrow\seq{}$, $\rt{}$).
In particular, $\rt{}$ arranges all the user's real-time interaction items within another specific window in chronological order of interaction time.

Fig.\ref{fig:gr-workflow}(a) demonstrates the aggregation: Compared with Fig.\ref{fig:traditional-recsys} where only one candidate is predicted, Fig.\ref{fig:gr-workflow}(a) aggregates three items in one sample, reusing the same user representation. Formally, it forms a feature representation for the same user: 
\vspace{-1ex}
\begin{eqnarray} \label{eq:user-aggregation-sample}
    \mathbb{D} = [\user{}, \overrightarrow\seq{}, \rt{}, [\cross{}, \itemfeature{}]_1, ..., [\cross{}, \itemfeature{}]_K]
\end{eqnarray}

By aggregating candidates into one sample, \method{} reduces a tremendous amount of resource by performing only a single computation and producing scores for all candidates.
% Besides, during training, user's request and corresponding candidates in one day is aggregated, leading to one feature representation comprised by one set of user-dependent features and multiple candidates' features. 
Specially, the user aggregation procedure greatly reduce the training samples from the sizes of all candidates to all users.
For inference, the candidates in a request are grouped as mentioned above, \method{} only perform one-time inference for all the candidate ranking, instead of inference of candidates' number. The aggregation circumvents the dependency of candidate' volume in inference cost, leaving possibility and potential for model scaling.

Eq.\ref{eq:user-aggregation-sample} is a combination of scalar and sequence features. 
To unify the input format, \method{} converts features and sequence into tokens.
Specially, for scalar features in $\user{}$, each feature is naturally converted to individual token with dimension $\Feat{}_{\user{}} \in \mathbb{R}^{N_{\user{}} \times d_{\uni{}}}$. $d_{\uni{}}$ is a set unified dimension for all tokens.
For sequence feature of $\overrightarrow\seq{}$ and $\rt{}$, each item $\seq{}$ is regarded as a token. Features in $\seq{}$ is firstly embedded and concatenated, then a MLP module is adopted for dimension unification. Formally, $\Feat{}_{\seq{}_i} = \text{MLP}(\text{Concat}(\Emb{}_{\iteminseq{}})) \in \mathbb{R}^{d_{\uni{}}}$.
% \begin{eqnarray}
%     \Feat{}_{\seq{}_i} = \text{MLP}(\text{Concat}(\Emb{}_{\iteminseq{}})) \in \mathbb{R}^{d_{\uni{}}}
% \end{eqnarray}
The features of $\seq{}$ in the sequence is concatenated along another dimension, leading to $\Feat{}_{\overrightarrow\seq{}} \in \mathbb{R}^{N_{\overrightarrow\seq{}} \times d_{\uni{}}}$.

Similar, each item $\itemfeature{}$ in candidates is regarded as a token. The features in candidates are embedded and concatenated, converted to the unified dimension by another MLP. Candidates are concatenated as a series of tokens: $\Feat{}_{\itemfeature{}} = \text{Concat}(\text{MLP}(\text{Concat}(\Emb{}_{\cross{}_i},\Emb{}_{\itemfeature{}_i}))) \in \mathbb{R}^{N_{\itemfeature{}}}$
% \begin{eqnarray}
%     \Feat{}_{\itemfeature{}} & = &  \text{Concat}(\Feat{}_{\itemfeature{}_i}) \nonumber \\
%     & = & \text{Concat}(\text{MLP}(\text{Concat}(\Emb{}_{\cross{}_i},\Emb{}_{\itemfeature{}_i}))) \nonumber \\ & \in &\mathbb{R}^{N_{\itemfeature{}} \times d_{\uni{}}}
% \end{eqnarray}
Finally, the constructed tokens from $\user{}$, $\overrightarrow\seq{}$, $[\cross{}, \iteminseq{}]$ are concatenated, leading a long sequence of tokens:
\begin{eqnarray}\label{eq:gr-feature}
    \Feat{}_{\mathbb{D}} & = &\text{Concat}([\Feat{}_{\user{}}, \Feat{}_{\overrightarrow\seq{}}, \Feat{}_{\rt{}}, \Feat{}_{\itemfeature{}}]) \\ \nonumber & \in & \mathbb{R}^{(N_{\user{}} + N_{\overrightarrow\seq{}} + N_{\rt{}} + N_{\itemfeature{}}) \times d_{\uni{}}}
\end{eqnarray}

\subsection{Unified HSTU Encoder}
Samples from a user are aggregated as a sequence of tokens, which is naturally suitable for processing using self-attention. 
Inspired by HSTU \cite{zhai2024actions}, \method{} uses stacks of self-attention layer and encoder-only architecture for modeling. 
% Specially, after embedding and token concatenation of Eq.\ref{eq:gr-feature}, though tokens are constructed from different domains (user profile, interaction history, candidate representation), a stack of self-attention module processes the input tokens in a unified way.

Similar in LLM, the input token sequences are processed layer-wisely. 
As shown in Fig.\ref{fig:gr-workflow}, in the self-attention block, input sequence $\mathbf{X}$ is firstly normalized by group layer norm. Features of the same domain (for example, $\user{}$) form a group. Then, the group layer norm ensures that tokens from different domains share a similar distribution before self-attention and aligns different semantic spaces of different domains $\tilde{X} = \text{GroupLN}(\X)$.
% \begin{eqnarray}
%     \tilde{X} & = & \text{GroupLN}(\X) \label{eq:input-layer-norm}
% \end{eqnarray}
The normalized input is then projected to 4 different representation: $\K, \Q, \V, \U = \text{MLP}_{\K/\Q/\V/\U}(\tilde{\X})$.
% \mathbf{U}$, $\mathbf{Q}$, $\mathbf{K}$, $\mathbf{V}$: 
% \begin{eqnarray}
%     \K, \Q, \V, \U & = & \text{MLP}_{\K/\Q/\V/\U}(\tilde{\X}) \label{eq:projection}
% \end{eqnarray}
$\mathbf{Q}$, $\mathbf{K}$ is utilized for multi-head attention computation with silu non-linear activation. The leading attention is divided by total length of input features as an average factor. Next, a customized mask ($\mathbf{M}$) is imposed on the attention score and the projected $\mathbf{V}$ is applied for value update:
\begin{eqnarray} \label{eq:attention}
    \tilde{\V} & = & \frac{\text{silu}(\K^T \Q)}{(N_{\user{}} + N_{\overrightarrow\seq{}} + N_{\rt{}} + N_{\itemfeature{}})} \mathbf{M} \V \label{eq:self-attention}
\end{eqnarray}
The projected $\mathbf{U}$ and the updated $\tilde{\V}$ is dot-producted. Then, another group layer norm is applied. Finally, we add a residual module and place another MLP above it:
\begin{eqnarray}\label{eq:dot-product}
    \X & = &  \text{MLP}(\text{GroupLN}(\tilde{\V} \odot \U)) + \X
\end{eqnarray}
% Similar to \cite{zhai2022scaling}, \method{} does not use FFN.

% \subsubsection{Dynamic Masking}
\textbf{Dynamic Masking}
\cite{zhai2024actions} utilizes the causal mask for sequence modeling. However, such a implementation does not bring in significant improvement in MTGR. 
% Besides, during training, user aggregation groups candidates from different requests, however, as $\rt{}$ records the nearest interaction of user, using simple auto-regressive mask for $\rt{}$ may lead to information leakage.
% during training, user aggregation groups candidates from different requests. 
Besides, since $\rt{}$ records the user's most recent interactions which timing might coincide with the sample aggregation window.
Using a simple causal mask in MTGR could result in information leakage.
For example, interactions in the evening should not be exposed to candidates in the afternoon, yet this information may be aggregated into one sample.
This dilemma requires a flexible and efficient masking.

In \method{}, $\user{}$, $\overrightarrow\seq{}$ is regarded as static (in the following, we denote $\user{}$, $\overrightarrow\seq{}$ as `static sequence') because its information comes from before the aggregation window, thus it does not cause causality errors. $\rt{}$ is dynamic as it progressively include user's interaction (`dynamic sequence' for $\overrightarrow\seq{}$) in real-time.  
$\method{}$ applies full attention for static sequence, auto-regressive with dynamic masking for $\rt{}$ and diagonal masking for inter-candidates. 
Specially, 3 rules are set for $\method{}$'s masking: 
% $\method{}$ applies dynamic masking, satisfying the following three rules:
\begin{itemize}
    \item The static sequence is visible to all tokens.
    % \item Dynamic sequence is visible to user tokens (static \& dynamic sequence) appear later.
    \item The visibility of the dynamic sequence adheres to causality, where each token is only visible to tokens that occur afterward, which include candidate tokens.
    \item Candidate tokens ($\cross{}$, $\itemfeature{}$) is visible to itself only.
\end{itemize}
Fig.\ref{fig:gr-workflow} (c) demonstrates an example for the dynamic masking:
`age', `ctr' represent feature token from $\user{}$; `seq1', `seq2' for $\overrightarrow\seq{}$; `rt1', `rt2' for $\rt{}$ and `target1' - `target3' for candidates. 
White block in row represents that the certain token is able to use the information from others while column represents the token is visible to others or not. 
% For example, `target1' is able to gather tokens for all except `target2'-`target3', which satisfy the rule No.2; Rule No.1 requires that tokens from `age' to `seq2' is visible to all other tokens. 
Full attention is utilized for $\user{}$ and $\seq{}$, leading to a white square from `age' to `seq2'.
For `rt1' to `rt2', we assume that `rt1' appear latter than `rt2', Therefore, a tiny square from `rt1' to `rt2' is constructed with white blocks in upper triangle, meaning that `rt1' is able to use information from `rt2' while `rt1' is not visible to `rt2'.
Besides, `target2' and `target3' is assumed to appear earlier than `rt1', leading to `rt1' is not visible to them. 
`rt2' appear earlier than all `target1' and `target2' while latter than `target3'. Therefore, `rt2' is not visible to `target3', which appears earlier than all `rt's, leading to its unable to use information from `rt's.

%% file: Contents/training.tex
To facilitate the design and development of \method~model structure and to conveniently incorporate more features from the fast-growing LLM community, we decided not to continue with our previous TensorFlow-based training framework. Instead, we opted to reconstruct our training framework within the PyTorch ecosystem. Specifically, we expanded and optimized TorchRec's functionality, made enhancements specifically catering to \method~model's characteristics, and ultimately accomplished efficient training of \method~model. Compared to TorchRec, our optimized framework improves training throughput by 1.6x – 2.4x while achieving good scalability when running over 100 GPUs. Compared to the DLRM baseline, we achieved 65x FLOPs per sample for forward computation, while the training cost remained nearly unchanged. We provide some of our key work as follows.

\textbf{Dynamic Hash Table.} TorchRec use a fixed-size table to process sparse embeddings, which is not suitable for large-scale industrial streaming training scenarios. Firstly, additional embeddings cannot be allocated to new users and items at real-time once the static table hits its preset capacity. 
% In cases where default embeddings are applied or eviction strategies are utilized to remove outdated IDs, model performance may suffer. 
Secondly, static embedding tables often require reserving more space than needed to avoid ID overflow, resulting in ineffective use of memory resources. 
To address these issues, we develop a high-performance embedding table utilizing hash techniques, enabling the dynamic allocation of space to accommodate sparse IDs during training. Our design employs a decoupled hash table architecture\cite{wang2025mtgrboost}, segregating key storage and value storage into separate entities. 
The key storage offers a lightweight mapping system linking keys to pointers, which point to embedding vectors, while the value storage contains the embedding vectors along with additional metadata, such as counters and timestamps, for eviction policies. 
This two-part system achieves two primary objectives: (1) facilitating dynamic expansion of capacity by replicating only the key storage instead of the extensive embeddings, and (2) improving the efficiency of key scanning by arranging keys in a compact format.

\textbf{Embedding Lookup.}
The embedding lookup process employs All-to-all communication for cross-device embedding exchange. To minimize duplicate ID transfers between devices, we implement a two-step process that ensures IDs remain unique both before and after communication.
% Furthermore, we have developed a feature configuration interface that enables automatic merging of tables, thereby decreasing the number of operators needed for embedding lookups and enhancing the overall efficiency of the process.

\textbf{Load balance.} 
In recommendation systems, user behavior sequences often exhibit a long-tail distribution where only a few users have long sequences, while most have short ones. This leads to significant computational load imbalance when training with a fixed batch size (abbreviated as BS). 
A common solution is to use sequence packing techniques\cite{krell2021efficient}, where multiple short sequences are merged into a longer one. However, this approach requires careful mask adjustments to prevent different sequences from interfering with each other during the attention calculation, which can be quite costly to implement. 
Our straightforward solution is to introduce dynamic BS. Each GPU's local BS is adjusted based on the actual sequence length of the input data, ensuring a similar computational load. Additionally, we've tweaked the gradient aggregation strategy to weight each GPU's gradients according to its BS, maintaining computational logic consistency with fixed BS.

\textbf{Other Optimizations.} To further enhance training efficiency, we implement pipeline technology utilizing three distinct streams: \textit{copy}, \textit{dispatch}, and \textit{compute}. 
The copy stream is responsible for transferring input data from the CPU to the GPU, while the dispatch stream executes table lookups using IDs, and the compute stream deals with both forward computations and backward updates. 
For example, while the compute stream performs forward and backward passes for batch $\mathbf{T}$, the copy stream simultaneously loads batch $\mathbf{T}+1$, thereby minimizing I/O delay. Once backward updates for batch $\mathbf{T}$ are completed, the dispatch stream promptly starts table lookups and communication for batch $\mathbf{T}+1$. Furthermore, we employ bf16 mixed-precision training and have designed a specialized attention kernel based on cutlass to accelerate training progress.

%% file: Contents/experiments.tex
\subsection{Experiment setup}
\textbf{Datasets.} 
Public datasets widely use independent ID and attribute features, where cross features are seldom introduced. However, cross features demonstrate importance in real-world application. Cross features are an important category of features in our scenario. They are often meticulously hand-crafted, including interactions like user-item, user and higher-level categories, item and spatio-temporal information.
To compensate the absence of cross feature in public datasets, we construct a training dataset based on logs from the real industrial-scale recommendation system in Meituan. Unlike public datasets, our real dataset contains a richer cross features set and longer user behavior sequences.
Using our industrial-scale data for experiments better highlights the significant impact of these cross features on real recommendation systems. 
% The significant impact of cross features is demonstrated in our industrial-scale data.
In addition, the volume of our dataset is large, allowing complex models to achieve more adequate convergence during training.
For the offline experiments, we collect data over a $10$-day period. The statistics of the dataset are shown in Table ~\ref{tab:dataset}. For the online experiments, in order to compare with the DLRM baseline that has been trained for over $2$ years, we constructed a longer-term dataset for the experiments, using data spanning more than $6$ months.

\begin{table}[htbp]
\vspace{-0.5em}
\captionsetup{font={small}}
\caption{Dataset Statistics}
\centering
\scalebox{0.83}{
\begin{tabular}{lccccc}
\toprule
Dataset & \#Users & \#Items & \#Exposure & \#Click & \#Purchases \\
\midrule
Train 
% & 205,985,588 
& 0.21 billion
& 4,302,391 
% & 23,741,406,845 
& 23.74 billion
% & 1,076,693,007 
& 1.08 billion
% & 175,736,671
& 0.18 billion
\\
Test & 3,021,198 & 3,141,997 & 76,855,608 & 4,545,386 & 769,534 \\
\bottomrule
\end{tabular}
}
\label{tab:dataset}
\vspace{-0.5em}
\end{table}

\textbf{Baseline.} 
For DLRM, we compare two methods in sequence modeling: SIM based on sequence retrieval and End2End modeling for original long sequences (E2E). In terms of scaling, we compared DNN, MoE\cite{ma2018modeling}, Wukong\cite{zhang2024wukong}, MultiEmbed\cite{guo2023embedding}, and UserTower.

MoE uses $4$ experts, with each expert containing a network of the same complexity as the base DNN. Wukong and MultiEmbed are configured to the same computational complexity as MoE. UserTower uses a set of learnable queries and inserts a qFormer\cite{li2023blip} layer with another MoE (16 experts) module on user behavior. UserTower's computational complexity is tripled than MoE method, but it can share this computation for multiple predicted items for the same user during inference, thereby reducing inference costs. It has achieved good results in our scenario.

$\method{}$ employs E2E to handle all sequence information. Additionally, as shown in Table~\ref{tab:setting}, we have set up three different scales to verify the scalability of $\method{}$.

\begin{table}[htbp]
\vspace{-0.5em}
\captionsetup{font={small}}
\caption{Comparison of models with different settings and computational complexity}
\centering
\scalebox{0.70}{
\begin{tabular}{lccc}
\toprule
Model & Setting & Learning rate & GFLOPs/example \\
\midrule
UserTower-SIM & - & $8 \times 10^{-4}$ & 0.86 \\
MTGR-small & $n_{\text{layer}}=3$, $d_{\text{model}}=512$, $n_{\text{heads}}=2$ & $3 \times 10^{-4}$ & 5.47 \\
MTGR-medium & $n_{\text{layer}}=5$, $d_{\text{model}}=768$, $n_{\text{heads}}=3$ & $3 \times 10^{-4}$ & 18.59 \\
MTGR-large & $n_{\text{layer}}=15$, $d_{\text{model}}=768$, $n_{\text{heads}}=3$ & $1 \times 10^{-4}$ & 55.76 \\
\bottomrule
\end{tabular}
}
\vspace{-1.0em}
\label{tab:setting}
\end{table}

\textbf{Evaluation Metrics.}
In offline, we focus on learning of two tasks: CTR and CTCVR (Click-Through Conversion Rate) and use AUC\cite{ferri2011coherent} and GAUC (Group AUC) for evaluation. 
GAUC is averaged on the the AUCs under users.
Compared to AUC, GAUC pays more attention to the model's ranking ability for the same user. 
For online evaluation, we focus on two metrics: PV\_CTR (CTR per page view) and UV\_CTCVR (CTCVR per user view), where UV\_CTCVR is the most crucial metric for evaluating the growth of business.

\textbf{Parameter Setting.}
Our model is trained using the Adam optimizer. For DLRM, each GPU processes a batch size of 2400, with 8 NVIDIA A100 GPUs for training. In the case of $\method{}$, the batch size is set to 96, utilizing 16 NVIDIA A100 GPUs for training. As shown in Table~\ref{tab:setting}, the learning rate decreases as the complexity of the model increases. Additionally, as the computational complexity grows, we proportionally enlarge the size of the sparse parameters by configuring different embedding dimensions.
Assuming that a token consists of $k$ features, the embedding dimension for each feature is generally set to an integer close to $d_{\uni{}}/k$. 
It is worth noting that, to prevent excessive overhead due to the overexpansion of sparse parameters, we primarily increase the dimensions of sparse features with smaller cardinality, while maintaining the dimensions of extremely sparse features unchanged.
Finally, the maximum length of $\overrightarrow{\seq{}}$ is set to $1000$, while $\rt{}$ is set to $100$.

\subsection{Overall Performance Comparison}
We evaluate the performance of MTGR and other baseline methods using our 10-day dataset.
Table~\ref{fig:performance} shows the performance of different models. The differences among various models across different offline metrics are quite consistent.
Based on previous experience, an increase $0.001$ in our offline metric is considered significant. 
Among the various versions of DLRM, Wukong-SIM and MultiEmbed-SIM achieve a better result than MoE-SIM. UserTower-SIM performs the best and UserTower-E2E shows a slight decline in performance compared to UserTower-SIM. We speculate that under the DLRM paradigm, the model complexity is insufficient to model all sequence information, leading to underfitting. 
% The three versions of $\method{}$ demonstrate good scalability, achieving better performance compared to DLRM.
Our proposed $\method{}$, even the smallest version, exceeds the strongest DLRM model. Additionally, models of three different sizes exhibit scalability, with their performance smoothly increasing as the complexity of the model increases.

\begin{table}[h]
\vspace{-0.5em}
\captionsetup{font={small}}
\caption{Overall performance. Impr.\% represents the relative improvement of the best MTGR model compared to the strongest DLRM baselines (underlined).}
\scalebox{1.0}{
\begin{tabular}{lcccc}
\toprule
     \multirow{2}{*}{Model} & \multicolumn{2}{c}{CTR} & \multicolumn{2}{c}{CTCVR} \\
    \cmidrule(lr){2-3} \cmidrule(lr){4-5}
     &  AUC & GAUC &  AUC & GAUC \\
    \midrule
DNN-SIM         & 0.7432  & 0.6679  & 0.8737  & 0.6504  \\ 
MoE-SIM         & 0.7484  & 0.6698  & 0.8750  & 0.6519  \\
MultiEmbed-SIM  & 0.7501  & 0.6715  & 0.8766  & 0.6525  \\ 
Wukong-SIM      & 0.7568  & 0.6759  & 0.8800  & 0.6530  \\ 
UserTower-SIM   &  \underline{0.7593}  & \underline{0.6792}  & 0.8815  & \underline{0.6550}  \\
UserTower-E2E   & 0.7576  & 0.6787  & \underline{0.8818}  & 0.6548  \\ 
\midrule
MTGR-small      & 0.7631  & 0.6826  & 0.8840  & 0.6603  \\ 
MTGR-medium     & 0.7645  & 0.6843  & 0.8849  & 0.6625  \\ 
\textbf{MTGR-large}      & \textbf{0.7661}  & \textbf{0.6865} & \textbf{0.8862}  & \textbf{0.6646} \\ 
\midrule
Impr.\% & \textbf{0.8956} & \textbf{1.0748} & \textbf{0.4990} & \textbf{1.4656} \\ 
\bottomrule
\end{tabular}
}
\vspace{-0.5em}
\label{fig:performance}
\end{table}

\begin{table}[h]
\captionsetup{font={small}}
\caption{Ablation studies of MTGR}
\scalebox{1.0}{
\begin{tabular}{lllll}
\toprule
     \multirow{2}{*}{Model} & \multicolumn{2}{c}{CTR} & \multicolumn{2}{c}{CTCVR} \\
    \cmidrule(lr){2-3} \cmidrule(lr){4-5}
     &  AUC & GAUC &  AUC & GAUC \\
    \midrule
MTGR-small      
& 0.7631  & 0.6826 & 0.8840  & 0.6603  \\ 
w/o cross features     
& 0.7495  & 0.6689 & 0.8736  & 0.6514  \\ 
w/o GLN                
& 0.7606  & 0.6809  & 0.8826  & 0.6585  \\
w/o dynamic mask       
& 0.7620 & 0.6810 & 0.8828 & 0.6587 \\ 
\bottomrule
\end{tabular}
}
\vspace{-0.5em}
\label{tab:ablation}
\end{table}

\subsection{Ablation Study}
We perform ablation studies on two components of $\method{}$: Dynamic Masking and group layer norm (GLN) based on our small one. The ablation results are shown in Fig~\ref{tab:ablation}.
Eliminating any of these from MTGR will result in a significant decline in performance, with the degree of decline being comparable to the increase from MTGR-small to MTGR-medium. This indicates the importance of Dynamic Masking and GLN for MTGR.
In addition, we conducted extra experiments on the importance of cross features to MTGR. After removing the cross features, the performance metrics show a significant drop, which even erases the gain of $\method{}$-large over DLRM, highlighting the vital role of cross features in the real recommendation system.

\subsection{Scalability}
\begin{figure}
\vspace{-0.5em}
    \centering
    \includegraphics[width=0.98\linewidth]{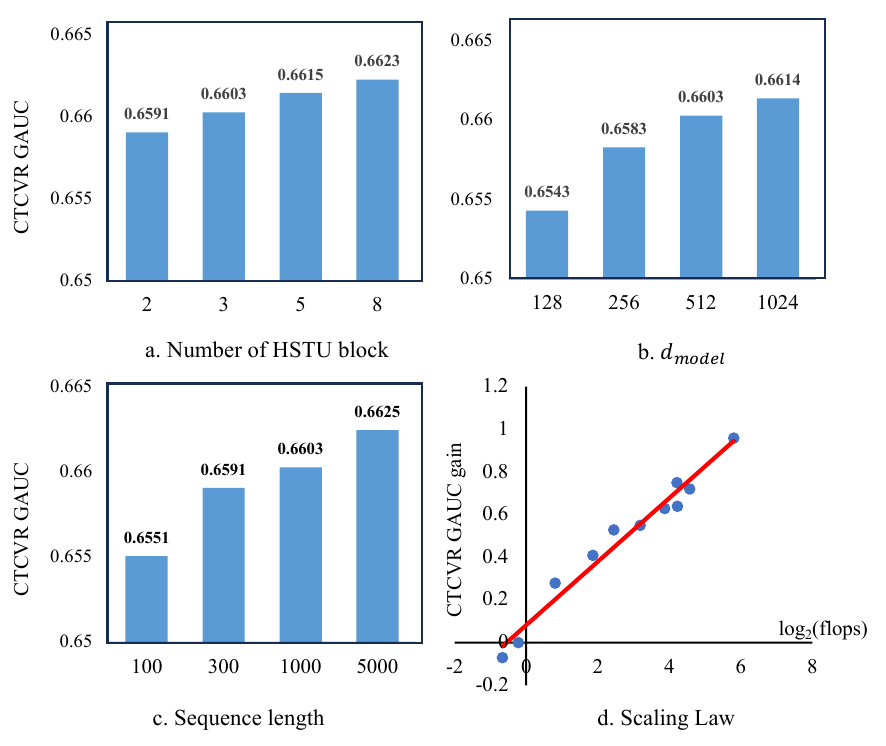}
    \captionsetup{font={small}}
    \caption{MTGR performance improves smoothly as we increase the numer of HSTU block, $d_{model}$ and sequence length for training}
    \vspace{-0.5em}
    \label{fig:scaling}
\end{figure}

Fig.~\ref{fig:scaling} illustrates the scalability of our $\method{}$. We perform tests based on MTGR-small for three different hyperparameters: the number of HSTU blocks, the $d_{\uni{}}$, and the input sequence length. As can be observed, MTGR demonstrates good scalability across different hyperparameters. Furthermore, Fig.~\ref{fig:scaling}(d) presents a power-law relationship between performance and computational complexity. The vertical axis indicates the gain in the CTCVR GAUC metric relative to our best DLRM model, UserTower-SIM, while the horizontal axis reflects the logarithmic multiple of computational complexity compared to UserTower-SIM.

\subsection{Online Experiments}
To further validate the effectiveness of $\method{}$, we deployed $\method{}$ in the Meituan Take-away platform, conducting AB testing with $2\%$ of the traffic. 
% This covers exposure in the tens of millions and hundreds of thousands of orders per day. 
The volume of experimental traffic covers millions of exposures per day, demonstrating the confidence of the experiment.
The comparison baseline is the most advanced online DLRM model (UserTower-SIM), and it has been continuously learning for $2$ years. We use data from the last $6$ months to train the $\method{}$ model, which is then deployed online for comparison. 

% We train small, medium, and large versions of the models. 
Although the volume of training data is significantly lower than the counterpart of DLRM model, the offline and online metrics still greatly exceed the DLRM base. As shown in Table ~\ref{tab:online_metrics}, both the offline and online metrics demonstrate scalability. We also found that as the number of training tokens increases, the benefits compared to DLRM continue to amplify. Eventually, in terms of CTCVR GAUC, our large version even exceeds the cumulative increase of all optimizations over the past year.

The model has already been fully deployed in our scenario, with training costs equal to DLRM, and inference costs reduced by $12\%$.
For DLRM, its inference cost is approximately linear with the number of candidates. However, $\method{}$ uses user aggregation for all candidates in a request, leading to sub-linear inference cost scaling with the number of candidates. This helps us reduce the overhead of online inference.

\begin{table}[h]
\captionsetup{font={small}}
\caption{Comparison of offline and online effects for different versions of MTGR.}
\vspace{-0.5em}
\scalebox{0.85}{
    \centering
    \begin{tabular}{lcccc}
    \toprule
     & \multicolumn{2}{c}{Offline Metric diff} & \multicolumn{2}{c}{Online Metric diff} \\
    \cmidrule(lr){2-3} \cmidrule(lr){4-5}
     & CTR GAUC & CTCVR GAUC  & PV\_CTR & UV\_CTCVR \\
    \midrule
    MTGR-small & +0.0036 & +0.0154 & +1.04\% & +0.04\% \\
    MTGR-medium & +0.0071 & +0.0182 & +2.29\% & +0.62\% \\
    MTGR-large & +0.0153 & +0.0288 & +1.90\% & +1.02\% \\
    \bottomrule
    \end{tabular}
    }
    \vspace{-0.5em}
    \label{tab:online_metrics}
\end{table}

%% file: Contents/conclusion.tex
% In the future, we will explore how supervised learning can be used to model knowledge transfer in domains without item or user overlap, to achieve broader applications of the explicit paradigm.
In this paper, we proposed MTGR, a new ranking framework to explore the scaling law in recommendation systems based on HSTU. MTGR combines the advantages of DLRM and GRM, allowing the use of cross-features to ensure model performance while having the same scalability as GRM. MTGR has already been deployed in our scenario and has yielded significant benefits.
In the future, we will explore how to extend MTGR to multi-scenario modeling, similar to large language models, to establish a recommendation foundation model with extensive knowledge.